\documentclass[3p,times,procedia]{elsarticle}
\usepackage{nupha_ecrc}

\volume{00}

\firstpage{1}

\journalname{Nuclear Physics A}

\runauth{}

\jid{nupha}

\jnltitlelogo{Nuclear Physics A}

\usepackage{amssymb}

\usepackage[figuresright]{rotating}

\newcommand{\ie}{{\it i.e.}}
\newcommand{\eg}{{\it e.g.}}

\begin{document}

\begin{frontmatter}



\dochead{}

\title{In-Medium Bottomonium Production in Heavy-Ion Collisions}


\author{Xiaojian Du$^1$, Min He$^2$, Ralf Rapp$^1$}
\address{$^1$Cyclotron Institute and Department of Physics \& Astronomy, Texas A\&M University, College Station, TX 77843-3366, USA\\
$^2$Department of Applied Physics,
Nanjing University of Science and Technology, Nanjing, China}

\begin{abstract}
We study bottomonium production in heavy-ion collisions using a transport model which utilizes kinetic-rate and Boltzmann equations to calculate the energy, centrality and transverse-momentum ($p_T$) dependence of the yields. Both gluo-dissociation and inelastic parton-induced break-up including interference effects are improved over previous work by using in-medium binding energies from a thermodynamic $T$-matrix approach. A coalescence model with bottom-quark spectra from Langevin simulations is implemented to account for thermal off-equilibrium effects in the $p_T$-spectra of the regeneration contribution. We also update the equation of state for the bulk medium by extracting it from lattice-QCD results. A systematic analysis of bottomonium observables is conducted in comparison to RHIC and LHC data. In particular, the off-equilibrium bottom-quark spectra are found to play an important role in the bottomonium $p_T$ spectra.
\end{abstract}
­

\end{frontmatter}


\section{Introduction}
\label{sec_intro}
Intense experimental~\cite{Adare:2014hje,Adamczyk:2013poh,Adamczyk:2016dzv,Abelev:2014nua,Fronze:2016gsr,Chatrchyan:2012lxa,Khachatryan:2016xxp,CMS:2016ayg} and theoretical~\cite{Grandchamp:2005yw,Emerick:2011xu,Strickland:2011aa,Song:2013tla,Zhou:2014hwa,Krouppa:2016jcl,Hoelck:2016tqf,Brambilla:2016wgg} efforts are underway to measure and interpret the systematics of bottomonium production in ultra-relativistic heavy-ion collisions (URHICs). Recent data on $\Upsilon$(1S), $\Upsilon$(2S) and $\Upsilon$(3S) states suggest a sequential suppression pattern. Unlike charmonia, where large contributions from regeneration have been established at LHC energies, bottomonium production could be dominated by the suppressed primordial contribution, which is supported by decreasing production yields with energy and centrality. However, the quantitative role of regeneration and its impact on the interpretation of current data remains an open issue.

In this paper we employ a rate equation approach for bottomonia, improved over previous work by the use of in-medium binding energies and extended to compute $p_T$ spectra (Sec.~\ref{sec_rate}), to calculate observables and compare them to the most recent data from RHIC and the LHC (Sec.~\ref{sec_rhiclhc}). We conclude in Sec.~\ref{sec_concl}.

\section{Transport model for bottomonia production}
\label{sec_rate}

The rate equation for bottomonium production in URHICs in the medium's rest frame~\cite{Grandchamp:2005yw},
\begin{equation}
\frac{\mathrm{d} N_Y(\tau)}{\mathrm{d}\tau} =
-\Gamma_Y(T)\left[N_Y(\tau)-N^{\rm eq}_Y(T)\right] \ ,
\end{equation}
involves two transport coefficients: the inelastic reaction rate, $\Gamma_Y$, and the thermal equilibrium limit $N^{\rm eq}_Y(T)$ for each state $Y=\Upsilon(1S), \Upsilon(2S), \chi_c, ...$.

\begin{figure}[t]
\includegraphics[width=0.48\textwidth]{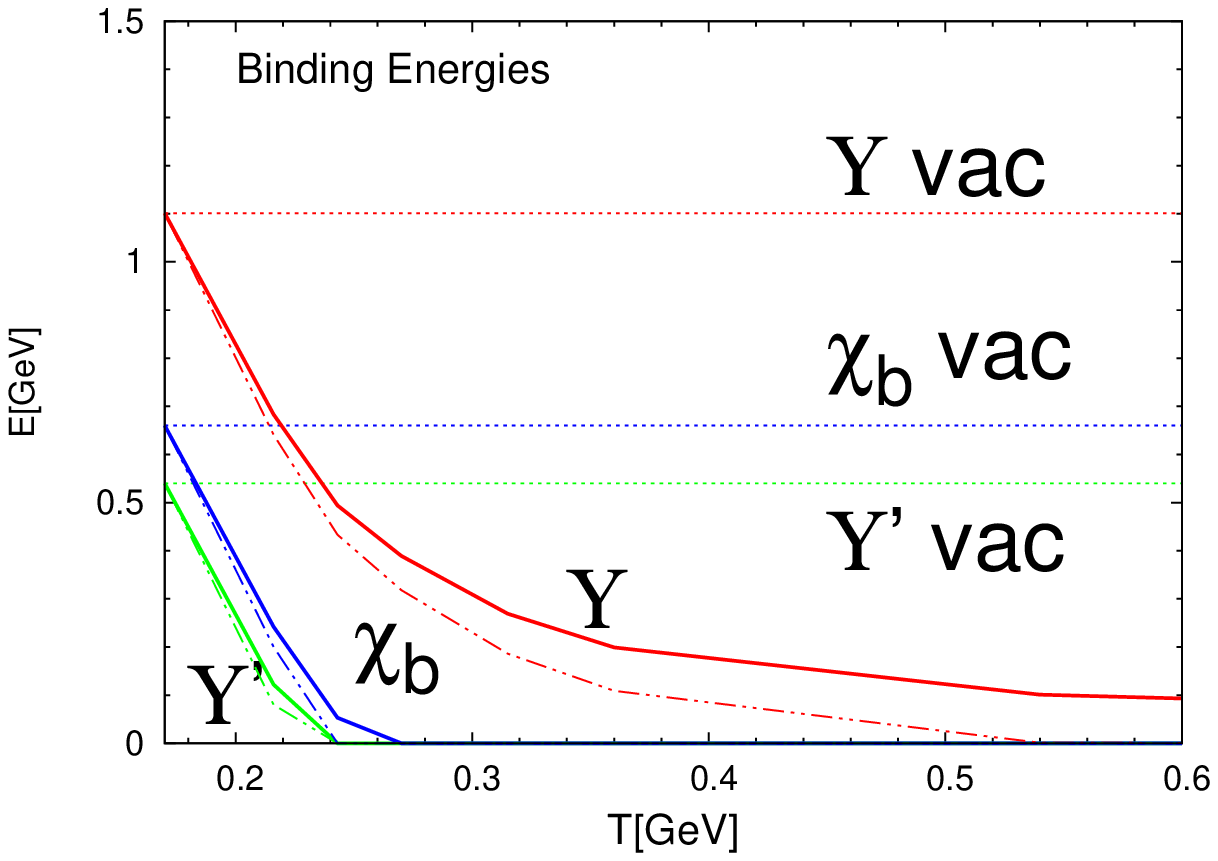}
\includegraphics[width=0.48\textwidth]{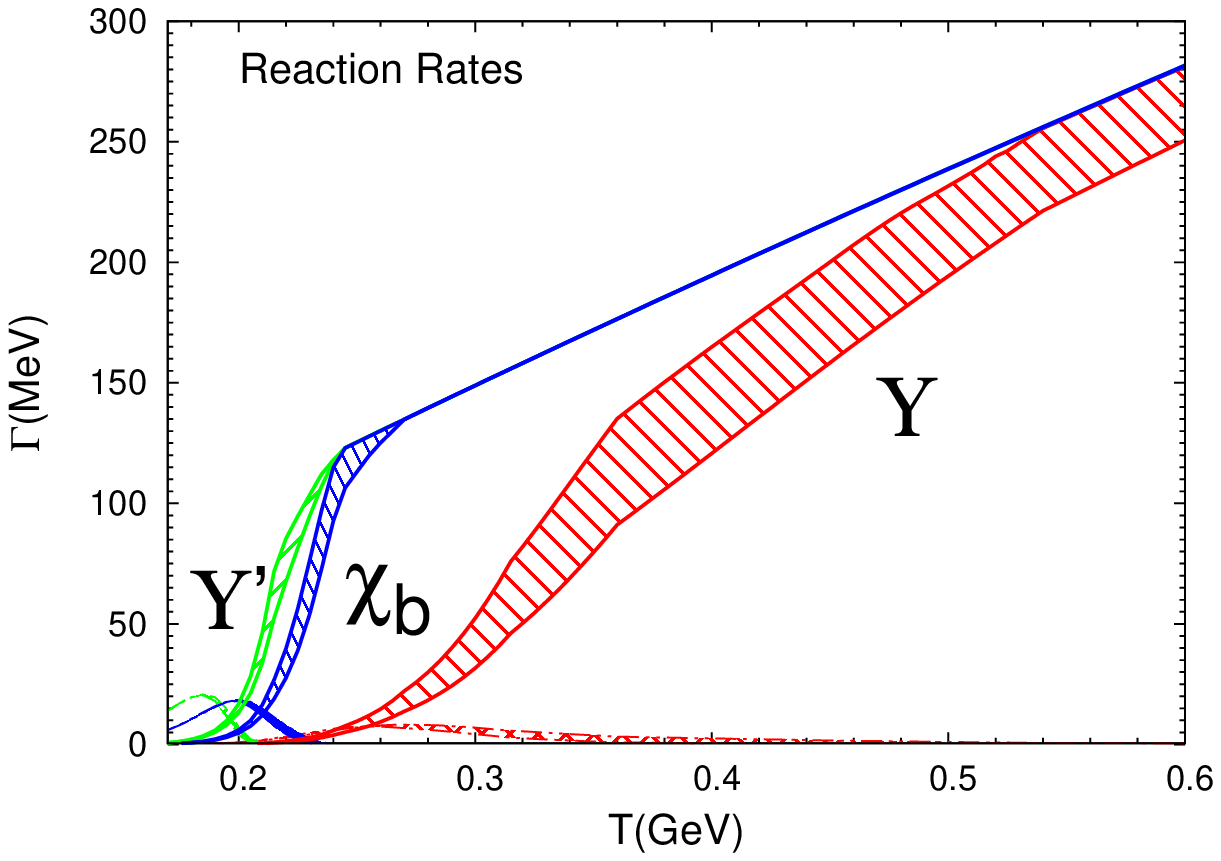}
\caption{Left panel: In-medium $Y$ binding energies extracted from $T$-matrix calculations~\cite{Riek:2010fk} (solid (dash-dotted) lines for $\eta$=1.0(1.1)), compared to their vacuum binding energies (dotted lines). The red, blue and green curves are for $\Upsilon(1S)$, $\chi_b$ and $\Upsilon(2S)$, respectively. Right panel: Inelastic reaction rates corresponding to the in-medium binding energies in the left panel with the band reflecting $\eta$=1.0-1.1. The solid (dash-dotted) line is the contribution from quasifree (gluo-) dissociation.}
\label{fig_bind}
\end{figure}
For the reation rates in the quark-gluon plasma (QGP) we include both gluo-dissociation and quasi-free mechanisms. The bottomonium binding energies strongly affect the reaction rates. We here employ the in-medium binding energy for $\Upsilon(1S)$ from microscopic $T$-matrix calculations (using the internal-energy potential)~\cite{Riek:2010fk}, and infer the ones for the excited states by assuming the in-medium $Y$ masses to be at their vacuum values (as suggested by the weak temperature dependence of lattice-QCD correlators~\cite{Aarts:2014cda}),
cf.~Fig~\ref{fig_bind} left. To account for pertinent theoretical uncertainties, we allow for deviations from the $T$-matrix results by defining a rescaling parameter $\eta$ so that $E_{\rm vac}-E_{\rm med}(T)=\eta(E_{\rm vac}-E_{T{\rm mat}}(T))$, where $\eta$$>$1 implies stronger medium effects, \ie, weaker binding. The reaction rates are dominated by the quasi-free mechanism; gluo-dissociation is only significant at temperatures where the total rate is small, cf.~right panel of Fig.~\ref{fig_bind}.

The thermal-equilibrium limit is evaluated in the QGP from the statistical model with bottom ($b$) quarks with a mass following from the procedure outlined above; a thermal relaxation factor is implemented to simulate incomplete
thermalization of $b$ quarks~\cite{Grandchamp:2002wp}. We neglect effects of the hadronic phase.

Using initial conditions obtained from data in $pp$ collisions  and potential cold-nuclear matter (CNM) effects (see below), the bottomonium yields are evolved with the rate equation in an expanding fireball background. Its temperature evolution is obtained from a fixed total entropy $S_{\rm tot}=s(T)V_{\rm FB}(t)$, adjusted to obtain the experimentally observed hadron yields at a chemical freezeout temperature of $T_{\rm c}$=170\,MeV. The nuclear modification factor is defined as a ratio of AA to $pp$ spectra, $R_{\rm AA}=(N^{\rm AA})/(N_{\rm coll}N^{\rm pp})$, scaled by the binary collision number, $N_{\rm coll}$, from the Glauber model.

The calculation of $p_T$ spectra employs the final yields of the primordial and regenerated $Y$'s from the rate equation. The Boltzmann equation is used to calculate the $p_T$-spectra of the primordial part via
\begin{equation}
\frac{\partial f(\vec{x},\vec{p},t)}{\partial t}+\frac{\vec{p}}{E_p}\cdot\frac{\partial f}{\partial \vec{x}}
=-\Gamma_Y(\vec{p},T)f(\vec{x},\vec{p},t)
\end{equation}
where $f(\vec{x},\vec{p},t)$ is the bottomonium phase space distribution and ${E_p}=\sqrt{p^2+m_Y^2}$ the $Y$ energy. The initial distribution $f(\vec{x},\vec{p},0)=f(\vec{x})f(\vec{p})$ is also calculated from the Glauber model for the spatial $N_{\rm coll}$ distribution, and from spectra in $pp$ collisions for the initial $p_T$ distribution.
For the regeneration component, incomplete $b$-quark thermalization mandates to go beyond our previously used blastwave approximation for charmonia. Instead we employ
$b$-quark spectra from Langevin simulations~\cite{He:2014cla} to
compute $Y$ $p_T$ spectra from a coalescence model~\cite{Greco:2003mm},
\begin{eqnarray}
\frac{\mathrm{d}^2N_{Y}(p_T,\phi)}{\mathrm{d}^2 p_{T}}=C_{\rm reg}
\int \mathrm{d}^2 p_{1t} \mathrm{d}^2 p_{2t} \frac{\mathrm{d}^2N_{b}}{\mathrm{d}^2 p_{1t}}
\frac{\mathrm{d}^2N_{\bar{b}}}{\mathrm{d}^2 p_{2t}} \
\delta^{(2)}(\vec{p}_{T}-\vec{p}_{1t}-\vec{p}_{2t}) \
\Theta \left[\Delta^{2}_{p}-
\frac{(\vec{p}_{1t}-\vec{p}_{2t})^2}{4}+
\frac{(m_{1t}-m_{2t})^2}{4}\right].
\label{coalescence}
\end{eqnarray}

\section{Bottomonia production at RHIC and LHC}
\label{sec_rhiclhc}
Our input cross section for bottom/onium production in $pp$ collisions at $\sqrt{s}$=0.2 and 2.76\,TeV are taken from our previous work~\cite{Emerick:2011xu}. For $\sqrt{s}$=5.02\,TeV we assume a 50\% increase relative to 2.76\,TeV. The feeddown on $\Upsilon$ states is updated to about 30(50)\,\% at low (high) $p_T$. For simplicity, we assume a 30\,\% feeddown from higher excited states (\eg, $3S$, $2P$, $3P$) to the $\Upsilon(2S)$ to be fully melted. The charged-particle density, $\frac{\mathrm{d}N_{ch}}{\mathrm{d}y}$, needed for the total entropy of the
fireball is increased by 22.5\,\% over
2.76\,TeV~\cite{Niemi:2015voa}.

\begin{figure}[t]
\includegraphics[width=0.48\textwidth]{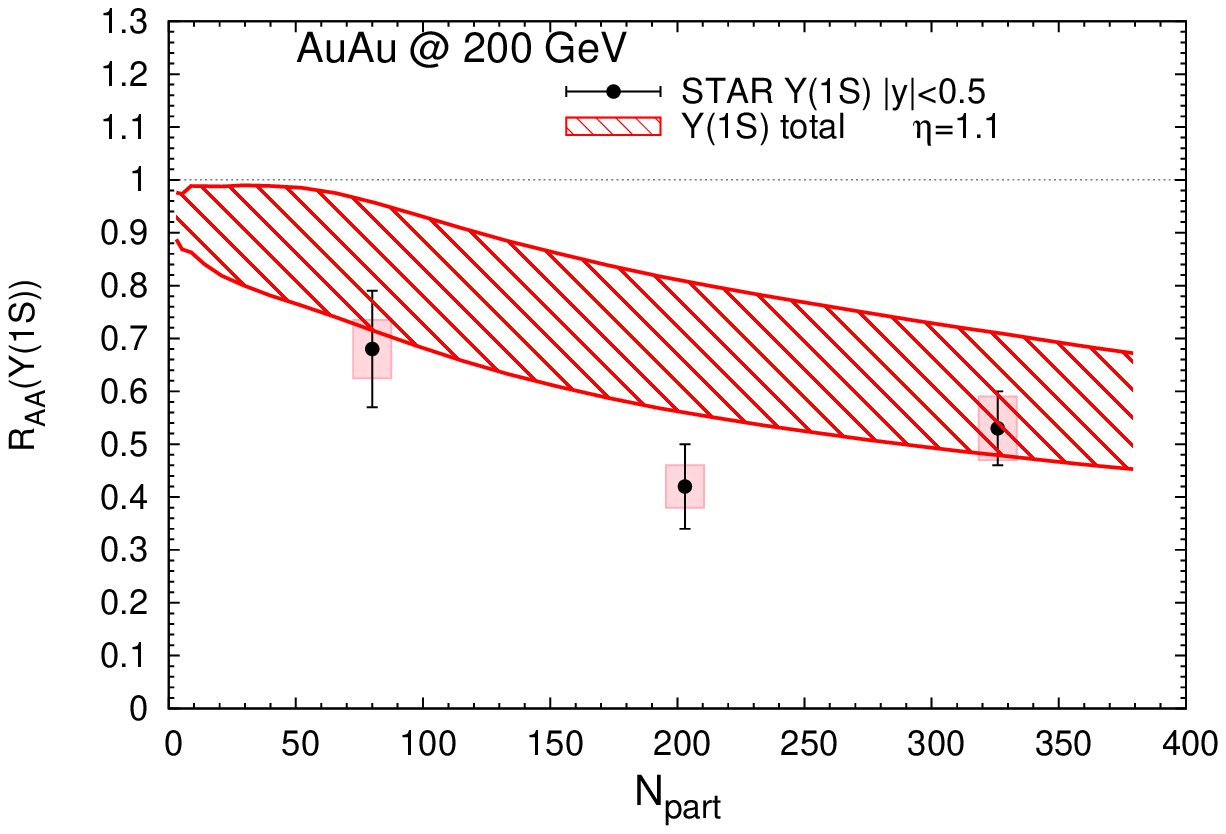}
\includegraphics[width=0.48\textwidth]{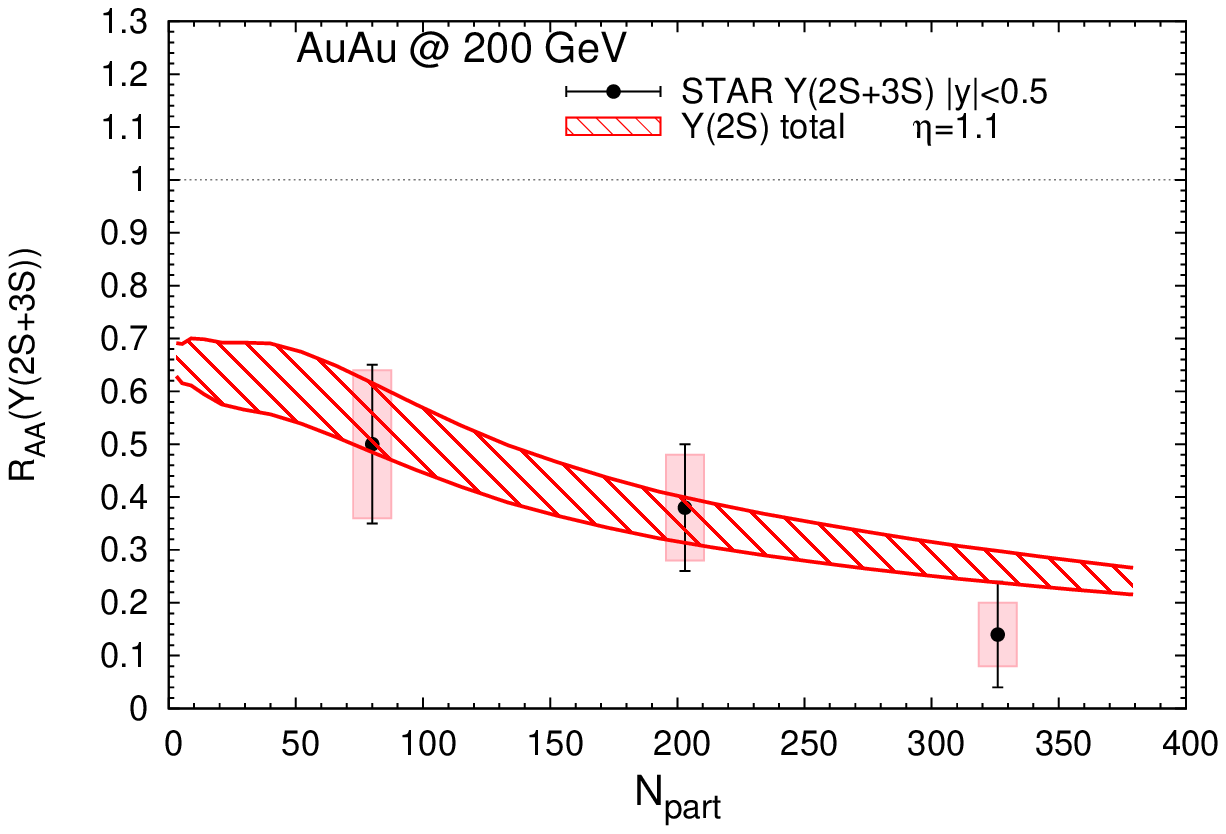}
\caption{Centrality dependence of the $R_{\rm AA}$ for
$\Upsilon(1S)$ (left) and $\Upsilon(2S)$ (right)
in 0.2\,TeV Au-Au collisions at RHIC, compared to
STAR data~\cite{Ye:2017qmstar}. The uncertainty bands are due to CNM effects represented by nuclear absorption cross sections, $\sigma_Y^{\rm abs}$=0-3\,mb.}
\label{fig_star}
\end{figure}

\begin{figure}[t]
\includegraphics[width=0.48\textwidth]{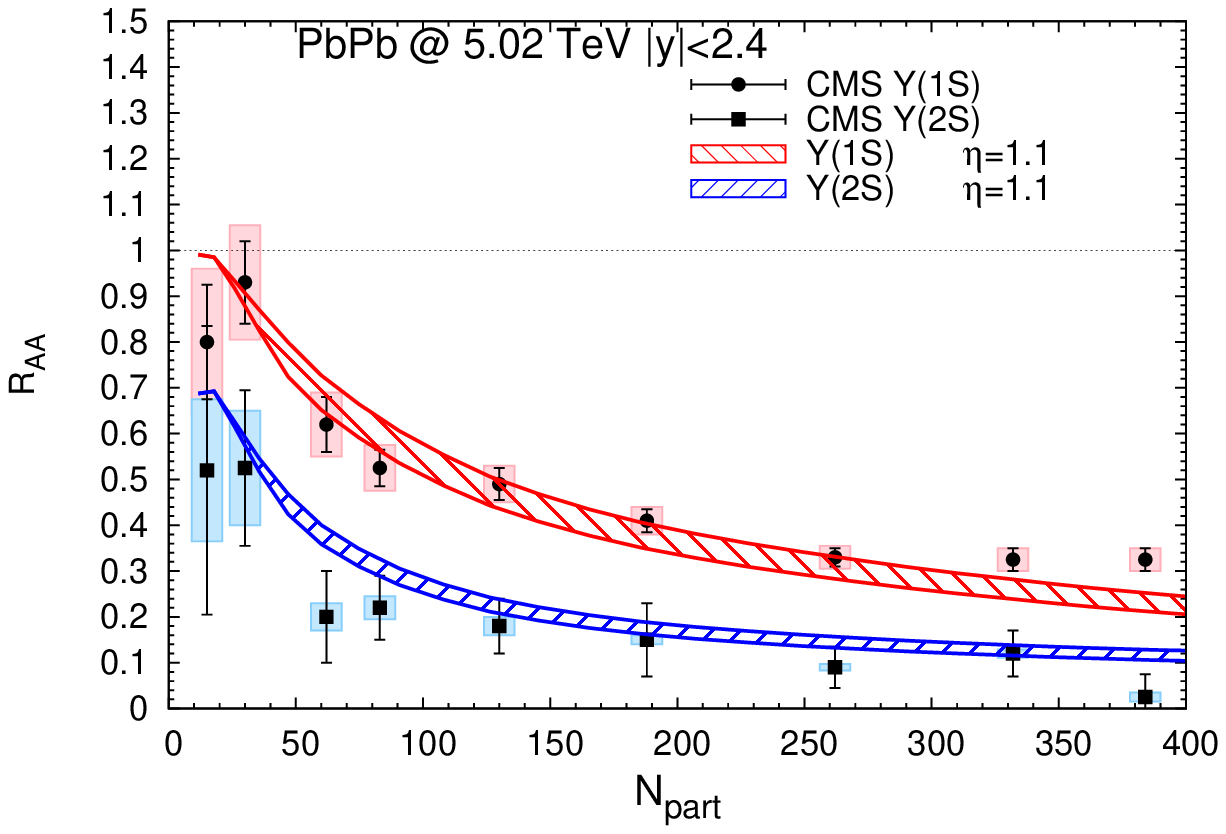}
\includegraphics[width=0.48\textwidth]{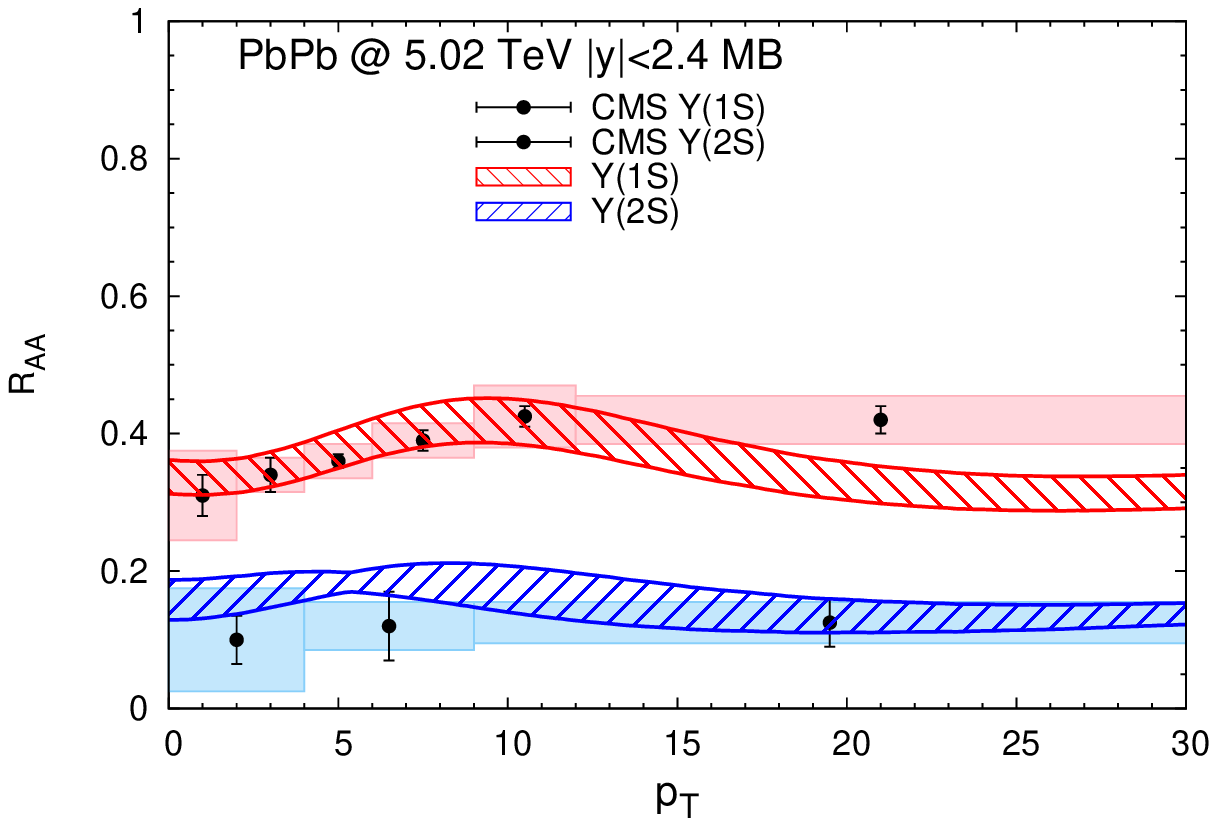}
\caption{Centrality (left) and transverse-momentum (right) dependence of the $R_{\rm AA}$ for $\Upsilon(1S)$ and $\Upsilon(2S)$ in 5.02\,TeV Pb-Pb collisions at the LHC, compared to CMS data~\cite{Flores:2017qmcms}. The bands represent a 0-15\,\% shadowing~\cite{Eskola:2009uj} on open-bottom and bottomonia.}
\label{fig_cms}
\end{figure}

In Figs.~\ref{fig_star} and \ref{fig_cms} we compare our predictions to the new mid-rapidity data presented by STAR (at
$\sqrt{s}$=0.2\,TeV)~\cite{Ye:2017qmstar} and CMS
(at $\sqrt{s}$=5.02\,TeV)~\cite{Flores:2017qmcms}, respectively, at this conference. The calculations have been carried out for the $\eta$=1.1 scenario which gave a slightly better agreement than $\eta$=1.0 in our previous comparison to CMS data at 2.76\,TeV.


Our predictions show a fair agreement with experimental data for the centrality dependence of both $\Upsilon(1S)$ and $\Upsilon(2S)$ at both collision energies. In particular, the rather strong suppression of the $\Upsilon(2S)$ observed by STAR
is accounted for (see right panel of Fig.~\ref{fig_star}).
In addition, the calculated $p_T$ spectra at 5.02\,TeV (see right panel in Fig.~\ref{fig_cms}) appear to capture the rather flat shapes in the CMS data. For the $\Upsilon(1S)$ the data show a hint for a slight rise for $p_T\lesssim10$\,GeV. In the calculations a similar trend is generated by the coalescence component which features a maximum structure around $p_T\simeq10$\,GeV. This structure is due to the radial flow of the coalescing $b$-quarks. However, it would be much more pronounced (and lead to a discrepancy with the data) if thermalized $b$-quark spectra were assumed instead of the transport $b$-quark spectra.
The comparison of our results to the newest forward-rapidity ALICE 5.02\,TeV data can be found in Ref.~\cite{Das:2017qmalice}; the agreement improves relative to 2.76\,TeV, where the ALICE data show significantly more suppression than our calculations.

\section{Conclusions}
\label{sec_concl}
We have calculated bottomonium production yields using a rate equation with in-medium transport coefficients in a QGP, augmented by a Boltzmann equation and coalescence model to compute $p_T$-spectra. The comparison of our predictions to
the newest experimental data released by STAR at 0.2\,TeV and by CMS and ALICE at 5.02\,TeV show fair agreement. Our calculations are compatible with appreciable reductions in the $Y$ binding energies, leading to a dissolution of the excited states at RHIC, while the ground-state $\Upsilon(1S)$ starts to dissolve at the LHC. The $\Upsilon(1S)$ suppression indeed shows a promising sensitivity to its in-medium binding energy, which can ultimately be used to determine more quantitatively the in-medium modifications of the QCD force, once other model uncertainties, such as the evolution of the fireball, CNM effects or the role of regeneration are controlled more precisely. At this point, regeneration plays a sub-dominant but significant role for the $\Upsilon(1S)$, possibly supported by the rising trend in the $p_T$ spectra of CMS at low $p_T$. Non-thermalized bottom-quark spectra turn out to play an essential role in the description of the $p_T$ spectra of the coalescence component. For the strongly suppressed $\Upsilon(2S)$ yields at the LHC, especially in central collisions, regeneration turns out to be the dominant production source. The slight over-prediction of our model for the centrality integrated $\Upsilon(2S)$ yields at both 2.76\,TeV and 5.02\,TeV might indicate a somewhat too large regeneration component. A potential resolution to this problem could be the emergence of $B$-meson states as $T_{\rm c}$ is approached from above, which leads to smaller $b$-quark fugacities and thus reduces the equilibrium limit of bottomonia in the regime near $T_{\rm c}$. Work in this direction is in progress~\cite{Du:2017DRM}.

\section*{Acknowledgement}
This work is supported by U.S National Science Foundation under grant no. PHY-1614484.













\end{document}